\documentclass[12pt,preprint]{aastex}

\shorttitle{Updated Big Bang Nucleosynthesis confronted to WMAP}
\shortauthors{A.~Coc, E.~Vangioni--Flam et al.}

\usepackage{graphicx}
\usepackage{dcolumn}
\usepackage{bm}

\newcommand{\zaa}{Astron.~Astrophys.}

\newcommand{\zapj}{Astrophys.~J.}

\newcommand{\zapjs}{Astrophys.~J.~S.}
\newcommand{\znas}{New~Astronomy}
\newcommand{\znat}{Nature}

\newcommand{\znp}{Nucl.~Phys.}

\newcommand{\zpr}{Phys.~Rev.}

\newcommand{\zadndt}{At. Data Nucl. Data Tables}

\newcommand{\gap}{\mathrel{ \rlap{\raise.5ex\hbox{$>$}}
                    {\lower.5ex\hbox{$\sim$}}  } }
\newcommand{\lap}{\mathrel{ \rlap{\raise.5ex\hbox{$<$}}
	            {\lower.5ex\hbox{$\sim$}}  } }

\newcommand{\obh}{$\Omega_bh^2$}
\newcommand{\lyma}{Lyman--$\alpha$}
\newcommand{\deu}{$D$}
\newcommand{\tro}{$^3He$}
\newcommand{\qua}{$^4He$}

\newcommand{\sep}{$^{7}Li$}

\newcommand{\dix}{$^{10}B$}

\newcommand{\hli}{$^4He$, $D$, $^3$He and $^{7}Li$}

\begin{document}

\title{Updated Big Bang Nucleosynthesis confronted to WMAP observations
 and to the Abundance of Light Elements}

\author{Alain Coc}
\affil{Centre de Spectrom\'etrie Nucl\'eaire et de Spectrom\'etrie
de Masse, IN2P3-CNRS and Universit\'e Paris Sud, B\^atiment 104,\\ 
91405 Orsay Campus, France}

\author{Elisabeth Vangioni-Flam}
\affil{Institut d'Astrophysique de Paris, CNRS, 98$^{bis}$ Bd Arago,
 75014 Paris, France}

\author{Pierre Descouvemont and Abderrahim Adahchour}
\affil{Physique nucl\'eaire Th\'eorique et Physique Math\'ematique, CP229,
 Universit\'e Libre de Bruxelles, B-1050 Brussels, Belgium}

\author{Carmen Angulo}
\affil{Centre de Recherches du Cyclotron, Universit\'e catholique de Louvain,
Chemin du Cyclotron 2, B-1348 Louvain--La--Neuve, Belgium}

\begin{abstract}
We improve Standard Big Bang Nucleosynthesis (SBBN) calculations taking into account
new nuclear physics analyses (Descouvemont et al. 2003).
Using a Monte--Carlo technique, we calculate the abundances of light
nuclei (\deu, \tro, \qua\ and \sep) versus the baryon to photon ratio.
The results concerning \obh\ are compared to relevant astrophysical 
and cosmological observations : 
the abundance determinations in primitive media and the
results from CMB experiments, especially the WMAP mission.
Consistency between WMAP, SBBN results and $D/H$ data
strengthens the deduced baryon density and has interesting 
consequences on cosmic chemical evolution. A significant discrepancy 
between the calculated \sep\ deduced from WMAP and the Spite plateau is clearly revealed. 
To explain this discrepancy three possibilities are invoked : systematic uncertainties
on the $Li$ abundance, surface alteration of $Li$ in the course of stellar evolution or 
poor knowledge of the reaction rates related to $^{7}Be$ destruction.
In particular, the possible role of the up to now neglected 
$^7$Be(d,p)2$\alpha$ and $^7$Be(d,$\alpha$)$^5$Li reactions is considered.
Another way to reconciliate these results coming from different horizons, consists to 
invoke, speculative, new primordial physics which could modify the nucleosynthesis 
emerging from the Big Bang and perhaps the CMB physics itself. 
The impressive advances in CMB observations provide a strong motivation for more 
efforts in experimental nuclear physics and high quality spectroscopy to keep 
BBN in pace.
 \end{abstract}


\newpage

\section{Introduction}

There exist different ways to determine the baryonic density of
the Universe. 
The "traditional method" is Standard
Big-Bang Nucleosynthesis (SBBN) which is based on nuclear physics in the early
universe. 
This calculation reproduces the primordial light
element (\deu, \tro, \qua\ and \sep)
abundances over an interval of 10 orders of magnitude.
Recently, however, the study of the Cosmic Microwave
Radiation (CMB) anisotropies and the census of
the \lyma\ forest at high redshift have provided new methods to obtain \obh.
In the case of the CMB, the baryonic parameter (\obh, where $h$
is the Hubble parameter expressed in units of 100 $km.s^{-1}.Mpc^{-1}$)  
is extracted from the amplitudes of the acoustic peaks in the angular power spectrum 
of the anisotropies. 
The \obh\ values deduced from these three different methods are in rather
good agreement but may not be totally model independent.

A series of data have been released by many experiments,
  but very recently, the WMAP mission has 
delivered a wealth of results, based on the first year of observations
(Spergel et al. 2003). 
The mean value \obh=0.024$\pm$0.001 agrees with the previous
estimates but the error bar is considerably reduced.
When including constraints from other observations at complementary 
angular scales, the value \obh=0.0224$\pm$0.0009 is obtained (Spergel et al. 2003),  
setting stringent constraints on the general discussion of the BBN scenario. 

In the case of the \lyma\ forest, the baryonic density is deduced from the study of the 
atomic HI and HeII \lyma\ absorption lines observed on the line of sight to quasars 
(baryonic matter distributed on large scales, in the redshift range 0$<z<$5).
Indeed, this evaluation, though indirect because of the relatively large ionization 
uncertainties, leads to results consistent with the two other methods 
(\obh$\sim$0.02, Riedeger et al. 1998).
However, the baryonic density obtained in this way carries a relatively large error bar, 
which in the present context makes it less constraining.

Consequently, due to the large efforts made recently to determine the 
cosmological parameters, it is now mandatory
to refine the BBN analysis. In this paper, we update the
study performed in Coc et al. (2002) where we had exploited a set of reaction rates 
from the NACRE compilation (Angulo et al. 1999). 
We reconsider here the BBN calculation using reaction rates obtained from a new analysis 
of the ten most important nuclear reactions (Descouvemont et al. 2003, hereafter DAA).
Moreover, we consider the impact on the BBN results of these main reactions and, at the
same time, study other reactions which could be potentially important for SBBN.

After a summary of the observational data concerning the light isotope
abundances and the new nuclear input, we use Monte--Carlo calculations 
to obtain the abundances of light nuclei (\deu, \tro, \qua\ and \sep) 
versus the baryon to photon ratio, taking into account the uncertainties on 
nuclear reaction rates.
We discuss both agreements and discrepancies confronting
calculations, abundance data and WMAP results.
 
\section{Abundances of light elements}
\label{s:obs}

The observation of the most primitive astrophysical sites in which abundances can be 
measured and their confrontation to the BBN calculations allow to extract \obh.  
For a general discussion on the updated observational data, see the review 
of Olive (2003).

The primordial \qua\ abundance, $Y_P$, is derived from observations of metal--poor, 
extragalactic, ionized hydrogen (HII) regions.

We adopt here the two recent values of Izotov et al. (1999), 
($Y_P = 0.2452\pm0.0015$) and Luridiana et al. (2003), 
($Y_P = 0.2391\pm0.0020$), giving a relatively large
range of abundance for this isotope.   
Indeed, when considering systematic uncertainties, 
Fields and Olive (1998) obtain the range $Y_P = 0.238 \pm0.002 \pm0.005$,

Deuterium is particularly fragile and is only destroyed in stellar processes. 
Hence, the primordial abundance should be represented, in principle, by the highest
value observed in remote cosmological clouds on the line of sight of high redshift 
quasars.
This is what we adopted in Coc et al. (2002) (hereafter CV).
However, recently, Kirkman et al. (2003) have obtained a new measurement of
$D/H = (2.42^{+0.35}_{-0.25})\times10^{-5}$ 
and $[O/H] = -2.79\pm0.05$.
They give also their best estimate of the primordial D abundance, averaging individual
measurements toward five QSOs,  namely $D/H = (2.78^{+0.44}_{-0.38})\times10^{-5}$ that we now adopt here.
However, as the sample of cosmological clouds is very limited and the systematic errors 
on $D/H$ values are hard to estimate, this value has to be considered with caution.
Indeed, Crighton et al. (2003) highlight important aspects of the analysis which were 
not explored in previous works showing that the methods used in analyses of $D/H$ in 
quasar spectra should be improved. 
For example, according to different hypotheses about contamination, 
 they show that $D/H$ in the absorber
toward QSO PG 1718+4807 can be as high as $4.2 \times 10^{-4}$ or 
significantly lower than $3.\times10^{-4}$.

Since the discovery of the Spite plateau (Spite and Spite 1982),
namely the constant lithium abundance as a function of
metallicity, many new observations have strengthened its existence.
Ryan et al. (1999, 2000) have obtained a tight limit on the plateau abundance.
Specifically, these authors take into account all possible
contributions from extrapolation to zero metallicity, \sep\ depletion mechanisms and 
biases in the analysis.
Their extrapolated value (at 95\% confidence level) is :
$Li/H$ = $(1.23^{+0.68}_{-0.32})\times10^{-10}$.
Recently, Th\'evenin et al. (2001) have obtained  VLT - UVES high resolution
spectra of seven metal poor stars in  the globular cluster, NGC 6397. 
Their mean value of lithium, A(Li)= 2.23$\pm$0.07, is consistent with the preceding one. 
Bonifacio et al. (2002) who have also observed this globular cluster, obtain a higher
mean value : A(Li) = 2.34 $\pm$0.056. The difference between these two evaluations
lies in the different effective temperatures adopted.
Indeed, these two independent observations and analyses give 
an indication of the systematic errors involved in $Li/H$ determination. 

Both observers and experts of stellar atmospheres agree to consider that the 
abundance determination in halo stars, and more particularly that of lithium require 
a sophisticated analysis . 
In this respect, the temperature scale is influential and it is possible
that the scale adopted by Ryan et al. (2000) underestimates the $Li/H$  ratio.
Moreover, the determination of $Li/H$ in stars embedded in globular clusters
is more questionable that in the halo field stars since globular cluster stars 
may be polluted by their environment. 
So, it would be necessary to select in a first step, star by star, those which are 
the less contaminated, i.e., the most adequate to give a reliable $Li/H$ abundance.
 Note, however, that stars from small globular clusters (as NGC 6397) are representative 
 of the halo stars (Cayrel, private communication).
In addition to the Ryan et al. range, adopted here as in CV, we will also consider, 
conservatively, 
the upper limit of the Bonifacio et al. (2002) value, namely $Li/H = 2.49\times10^{-10}$.
Note however that these globular cluster determinations, at [Fe/H]$\approx$-2, 
cannot be directly compared to the Ryan et al. (2000) extrapolated value.

$^3He$ has been measured recently by Bania et al. (2002) in HII regions, but due to 
the large scatter in the data and the complex galactic history of this isotope, we
cannot consider it as a good cosmological tracer (Vangioni-Flam et al. 2003).

\section{SBBN with improved nuclear input}
\label{s:nucl}

In our previous work (CV), we performed Monte-Carlo 
calculations to obtain statistical limits on the calculated abundances, using 
mainly the NACRE compilation of reaction rates  (Angulo et al. 1999).
One of the main innovative features of NACRE with respect to former
compilations (Caughlan and Fowler 1988, hereafter CF88) is that  uncertainties 
are analyzed in detail and
realistic lower and upper bounds for the rates are provided.
However, since it is a general compilation for multiple applications, 
coping with a broad range of nuclear configurations, 
these bounds have not always been evaluated through a rigorous 
statistical methodology.
Hence, in CV, a simple uniform distribution between these 
bounds was assumed for the Monte--Carlo calculations. 
Since this 
compilation was not specifically addressed to the nuclear reactions implied in 
the BBN it had also to be complemented by other sources (Smith et al. 1993, 
Brune et al. 1999).
Two recent SBBN calculations have been made with updated reaction rates, 
one based on the Nollett and Burles (2000) compilation (hereafter NB)
and another (Cyburt et al. 2001; hereafter CFO) on a partial, reanalysis the 
NACRE compiled data.
These works (NB and CFO) have given better defined statistical limits for the 
reaction rates of interest for SBBN. 
One (NB) has used spline functions to fit the astrophysical $S$--factors 
(see definition in CV) while the other (CFO) have used the NACRE 
$S$--factors with a different normalization (restricted to BBN energies.)
In NACRE, data are in general fitted either by Breit--Wigner formula 
(the shape of nuclear resonances) or by low order polynomial
for non-resonant contributions.    
Indeed, unlike in CFO,  the fits are not restricted to the energy range of 
BBN, taking advantage of all data to constrain the nuclear factor. 
The use of low order polynomials, or better theoretical $S$--factors shapes, 
rather than e.g. splines, has the advantage of smoothing out the dispersion of 
data arising from the measurement technique itself rather than from 
physics when no sharp resonance is expected in the energy domain.
Consequently, the CFO global normalization factors are different from 
those of NACRE.
One should note, however, the isotope yields obtained from BBN calculations 
using the two compilations (NACRE and NB) agree well, reinforcing the 
confidence in these analyses.

Nevertheless, in order to improve on the general NACRE compilation,
DAA have reassessed carefully the main nuclear network (ten reactions) on the basis of an 
R-Matrix analysis.
The R--matrix theory has been used for many decades in the nuclear physics
community. It allows to parametrize nuclear cross sections 
with a reduced set of parameters related to nuclear quantities such as resonance 
energies and partial widths. 
This method can be used for both resonant and non-resonant 
contributions to the cross section. (See DAA
and reference therein for details of the method.) 
The energy dependence of the fitted S-factors is now constrained by
the Coulomb functions and R-matrix poles, rather than by arbitrary
polynomial or spline functions. 
Even though this method has been widely used in nuclear astrophysics (see e.g. 
Barker \& Kajino 1991 for a recent application to a nuclear astrophysics problem), 
this is the first time that it is applied to SBBN reactions. 
In addition, this new compilation (DAA) provides 
1--$\sigma$ statistical limits for each of the 10 rates: 
$^2$H(p,$\gamma)^3$He, $^2$H(d,n)$^3$He, $^2$H(d,p)$^3$H, $^3$H(d,n)$^4$He
$^3$H($\alpha,\gamma)^7$Li, $^3$He(n,p)$^3$H, $^3$He(d,p)$^4$He, 
$^3$He($\alpha,\gamma)^7$Be, $^7$Li(p,$\alpha)^4$He and $^7$Be(n,p)$^7$Li.
These rate limits are derived from the R--Matrix parameter errors calculated 
during the fitting procedure (see DAA). 
The two remaining reactions of importance, n$\leftrightarrow$p and 
$^1$H(n,$\gamma)^2$H (Chen and Savage 1999) come from theory and are 
unchanged with respect to CV.

\begin{figure}
\begin{center}
\includegraphics[width=8cm]{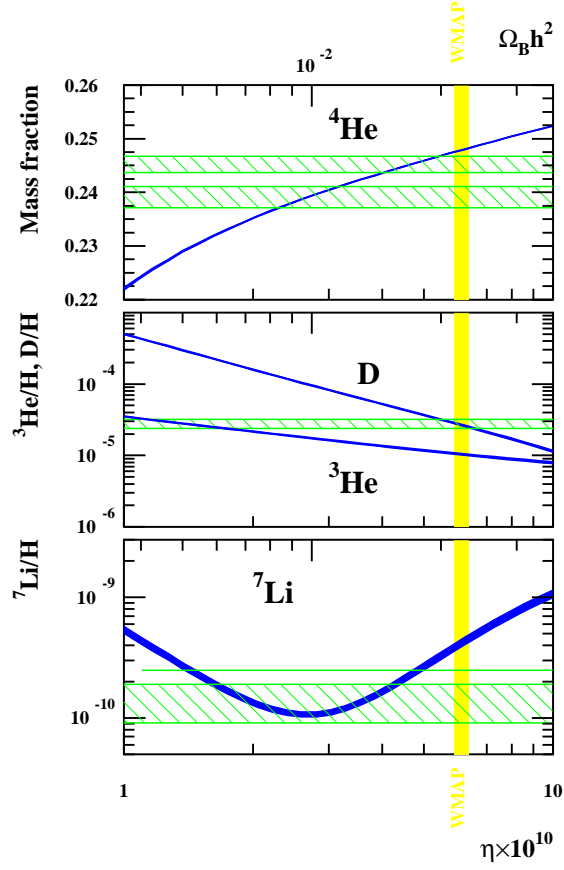}
\caption{
Abundances of \qua\ (mass fraction), \deu, \tro\ and \sep\ (by
number relative to H) as a function of the baryon over photon ratio 
$\eta$ or \obh\ .  Limits (1-$\sigma$) are obtained from Monte Carlo calculations. 
Hatched area represent primordial \qua, \deu\ and \sep\ abundances
deduced from different primitive astrophysical sites (see 
Section~\protect\ref{s:obs}): 
Izotov et al. (1999) (high area) and Luridiana et al. (2003) (low area) for \qua, 
Kirkman et al. (2003) for \deu, 
and Ryan et al. (2000) for \sep\ (95\% c.l.).
Concerning \sep, we also show an upper limit derived from
Bonifacio et al. (2002) observations (dashed line). 
The vertical stripe represents the (1-$\sigma$) 
\obh\ limits provided by WMAP (Spergel et al. 2003).
}
\label{f:heli}
\end{center}
\end{figure}

We have re--done our Monte-Carlo calculations using this time Gaussian distributions with 
parameters provided by the new compilation (DAA) discussed above. 
We have calculated the mean and the variance of the \qua, \deu, \tro\ and \sep\ yields 
as a function of $\eta$, fully consistent with our previous analysis (CV).   
The differences with CFO for the \sep\ yield is probably due to their 
renormalization procedure of NACRE S--factors.
Figure~\ref{f:heli} displays the resulting abundance limits (1-$\sigma$) 
[it was 2-$\sigma$ in Fig.4 of CV] from SBBN calculations compared to primordial 
ones inferred from observations. 
It is important to note that the present
 results are in good agreement with CV.
With these improved calculations, we can now compare SBBN results, primitive
abundances of the light elements and baryonic density derived from CMB observations.

\section{Discussion}

Following numerous determinations of \obh\ through CMB observations, WMAP 
observations and subsequent analyses, including other observational constraints,
have delivered a very precise value, \obh = 0.0224 $\pm$ 0.0009,
 corresponding to $\eta = (6.14\pm 0.25)\times10^{-10}$
(Spergel et al. 2003). 
In their paper, this evaluation has been compared to the BBN calculations of 
Burles et al. (2001), leading to $D/H = (2.62^{+0.18}_{-0.2})\times10^{-5}$. 
With our improved analysis of SBBN reaction rates, using
the WMAP \obh\ range together with these SBBN results (WMAP+SBBN 
hereafter), we can also deduce the primordial abundances as shown in
Figure~\ref{f:heli} where is represented the WMAP \obh\ range intercepting
the SBBN yield curves.  
The uncertainties on these abundances take into account the WMAP \obh\ 
uncertainty and the SBBN uncertainties from DAA reaction rates.  
Our WMAP+SBBN  deuterium primordial abundance is
$D/H$ = $(2.60^{+0.19}_{-0.17})\times10^{-5}$
which is in perfect agreement with the average value 
$(2.78^{+0.44}_{-0.38})\times10^{-5}$ (Kirkman et al. 2003) of $D/H$ observations in 
cosmological clouds. 
The other primordial abundances deduced from WMAP+SBBN 
are $Y_P$ = 0.2479$\pm0.0004$ for the \qua\ mass fraction, 
$^3$He$/H$ = $(1.04\pm0.04)\times10^{-5}$ and 
$^7$Li$/H$ = $(4.15^{+0.49}_{-0.45})\times10^{-10}$.
Recently, Cyburt et al. (2003) have also compared BBN and WMAP data. 
Their mean $D/H$ value ($2.75\times10^{-5}$) is slightly higher than our result while 
$Y_P$ is in good agreement. 
More important, their predicted \sep\ ($3.82\times10^{-10}$) is lower than 
our prediction (about 11\%; see Table~\ref{t:yields}). 
The reason is probably due to the different normalization for nuclear data 
as discussed above.
It is timely to confront these primordial nucleosynthesis 
results with the observations described in Section~\ref{s:obs} and to explore 
various astrophysical consequences.

\begin{table}
\caption{BBN results at WMAP \obh\ from different authors.}
\begin{flushleft}
\begin{tabular}{l|c|c|c|c}
\hline
Source&$Y_p$&$D/H$&$^3He/H$&$Li/H$\\
&&$\times10^{-5}$&$\times10^{-5}$&$\times10^{-10}$\\
\hline
This work& 0.2479$\pm0.0004$&$2.60^{+0.19}_{-0.17}$&$1.04\pm0.04$&
$4.15^{+0.49}_{-0.45}$\\
\hline
Cyb03 & $0.2484^{+0.0004}_{-0.0005}$
& $2.75^{+0.24}_{-0.19}$ & $0.93^{+0.055}_{-0.054}$ & $3.82^{+0.73}_{-0.60}$\\
\hline
Bur01 & - &$2.62^{+0.18}_{-0.22}$& -&-\\
\hline
\end{tabular}
\end{flushleft}
Cyb03: Cyburt et al. 2003; Bur01 : Burles et al. 2001\\
\label{t:yields}
\end{table}

\subsection{Helium}

As said previously, the \qua\  abundance determinations in HII regions are quite 
unsatisfactory due to observational uncertainties and the complex physics of
HII regions. 
Luridiana et al. (2003) obtained a new determination of $Y_P$, 
based on the abundance analysis of five metal poor extragalactic HII regions. 
This relatively low value ($0.2391\pm0.002$) differs significantly from
the Isotov et al. (1999) higher value ($0.2452\pm0.0015)$ and the one 
deduced from BBN+WMAP (0.2479), but systematic uncertainties may prevail 
due to observational difficulties and complex physics 
(Fields and Olive, 1998). 
In fact, the Izotov et al.
interval (Section~\ref{s:obs} and Fig.~\ref{f:heli}) 
is only marginally compatible with the WMAP observations.
($\sim$8\% probability).

\begin{figure}[htb]
\includegraphics[width=9cm]{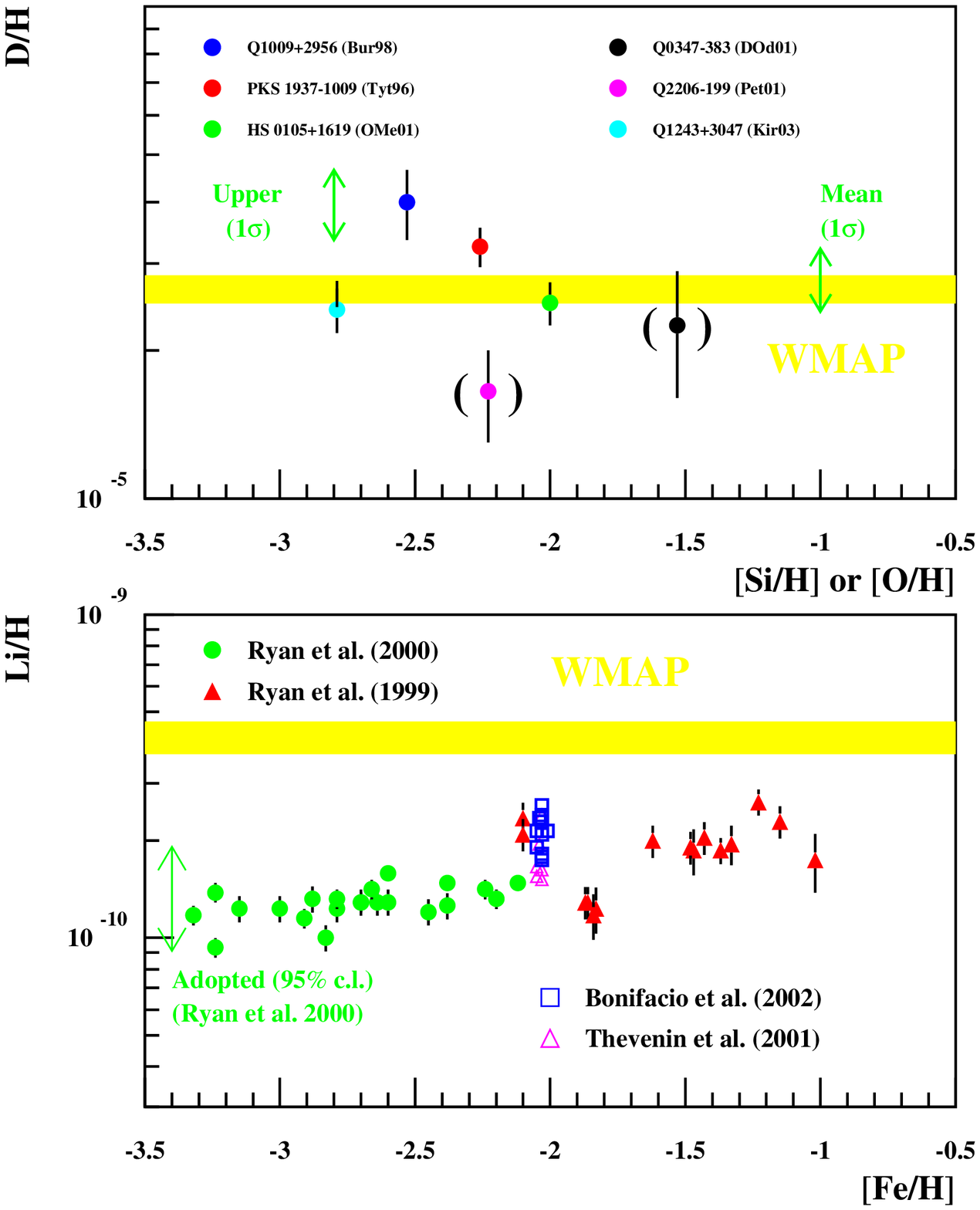}
\caption{Observed abundances as a function of metallicity from objects 
which are expected to reflect primordial abundances.
Upper panel : observed \deu\ abundances in cosmological clouds
(parenthesis indicate less established observations.)
The mean observational value (Kirkman et al. 2003) and the highest observed 
value used in CV are shown by arrows.  The horizontal stripe
 represents the (1-$\sigma)$ \obh\ limits provided by WMAP+BBN.
Lower panel : observed \sep\ abundances from Ryan et al. (1999; 2000) and 
extrapolated primordial abundance  Ryan et al. (2000) shown by an arrow.
$Li/H$ observations in a globular cluster at [Fe/H]=-2  
( Th\'evenin et al. 2001; Bonifacio et al. 2002) are also displayed. 
The horizontal stripe
 represents the (1-$\sigma)$ \obh\ limits provided by WMAP+BBN.
}
\label{f:obs}
\end{figure}

\subsection{Deuterium}

WMAP observations together with our SBBN calculations lead to
the mean primordial $D/H$ value of $2.60\times10^{-5}$. 
In Fig.~\ref{f:obs} we plot $D/H$ observations at high redshift 
(Burles \& Tytler, 1998; Tytler et al. 1996; O'Meara et al. 2001, 
D'Odorico et al. 2001; Pettini \& Bowen 2001; Kirkman et al. 2003)
which are thought to be representative of the \deu\ primordial 
abundances together with those
inferred from SBBN calculations and \obh\ range from WMAP.
The stripe widths represent the uncertainty (1-$\sigma$) originating from 
both the WMAP \obh\ and nuclear uncertainties. 
It shows that this result is consistent with $D/H$ observations at high 
redshift and specifically with the last measurement and averaged value
of Kirkman et al. (2003). 
The convergence between these two independent methods seems to confirm 
this \obh\ evaluation. 
Adopting this result as a firm basis, one can draw some consequences 
on the cosmic chemical evolution and on the global star formation rate history 
in the Universe. 
In addition to the high redshift data, the only $D/H$ observations available are 
i) the protosolar value which is affected by a large error bar,
$(2.5\pm0.5)\times10^{-5}$ (Hersant et al. 2001), and 
ii) the local and present value in the interstellar medium, 
$(1.52\pm0.08)\times10^{-5}$ (Moos et al. 2002).
Accordingly, these observations can only set constraints on the chemical evolution 
of our Galaxy showing that the star formation history is probably modest and smooth. 
It is worth noting that, in this context, \deu\ has almost not been
depleted between Big Bang and the sun birth, typically evolving from 
$2.60\times10^{-5}$ to $2.5\times10^{-5}$, during about 10 Gyr, whereas during 
the last 4.6 Gyr, the mean $D/H$ has decreased from $2.5\times10^{-5}$
to $1.5\times10^{-5}$.
This could seem paradoxical but, taking into account a possible primordial infall, 
one could alleviate  the problem of the proximity between the SBBN and present 
$D/H$ ratio (Chiappini et al. 2003).

On the other hand, the accumulation of information on the high redshift Universe 
leads to the conclusion that there was an intense activity in the past compared to 
present ($z$ = 0). 
Indeed, the cosmic star formation appears to be much higher at high $z$ 
(Lanzetta et al. 2002, Hernquist and Springel 2003) and moreover, many clues point 
toward the existence of an early generation of massive stars (Silk 2003, Cen 2003). 
In this case, the parameters governing global galactic evolution (initial mass function 
IMF, star formation rate SFR,..) should be reconsidered (see Scully et al. 1997,
 Daigne et al. 2003, in preparation).
In this context, the local D abundance is only
representative of local interstellar medium and not of the general star formation history of our Galaxy and a fortiori of 
the whole Universe. 
 All the more so the FUSE mission has revealed a complex landscape on the D abundance
within regions in the solar neighborhood. 
Indeed, although Moos et al. (2002) did not find any noticeable D variation within 100 pc
(local bubble), Hoopes et al. (2003)  find a $D/H$ ratio of less than $10^{-5}$ on longer lines of sight 
(a few hundreds pc). 
A third observed line of sight leads to an even lower D abundance, 
$D/H=0.52\pm0.09\times10^{-5}$ (H\'ebrard and Moos, 2003). 
Finally, these new results show clearly that it is dangerous to take as a reference any 
local value of $D/H$ without considering the systematic errors in the determination of 
the H column densities (Vidal-Madjar and Ferlet 2002). 
Starting from the primordial $D/H$ deduced from BBN+WMAP, one can predict, according to 
specific SFR histories versus $z$ (which are probably highly variable from one type of galaxy
to the other, see Kauffmann et al. 2003) very different present
 D abundances in spiral, elliptical galaxies... (Daigne et al. 2003, in preparation).

\subsection{Lithium}

Contrary to Deuterium, Lithium presents a neat discrepancy. Indeed, our value
deduced from WMAP+SBBN is \sep = $(4.15^{+0.49}_{-0.45})\times10^{-10}$, 
while the most recent observations of Lithium in halo stars lead to the range
$Li/H$ = $(1.23^{+0.68}_{-0.32})\times10^{-10}$ (95\% c.l.)
(Ryan et al. 2000). 
Hence, this observed $Li/H$ is a factor of 3.4 lower than the WMAP+SBBN value. 
Even when considering the corresponding uncertainties, the two $Li/H$ 
values differ statistically ($\sim3\times10^{-7}$ probability).
This confirm our (CV) and other (CFO, Cyburt et al. 2003) previous 
conclusions that the \obh\ range deduced from SBBN of \sep\ are only 
marginally compatible with those from the CMB observations available by 
this time.
Considering the different nuclear reaction rate analyses involved 
(NACRE, NB, CFO and DAA) this result is robust with respect to nuclear
uncertainties concerning the main SBBN reactions.
It is strange that the major discrepancy affects \sep\ since 
it could a priori lead to more reliable primordial value than deuterium, 
because of much higher observational statistics and an easier extrapolation 
to primordial values.
In  Fig.~\ref{f:obs} (lower panel) are shown the most recent \sep\ 
observations by Ryan et al. (1999, 2000) as a function of metallicity 
for old halo stars together with their extrapolated primordial 
$Li/H$. The data of Th\'evenin et al. (2001) and Bonifacio et al. (2002) are also 
included.
This figure emphasizes a strong incompatibility between WMAP+SBBN 
 and measurements made in halo stars. 
This large difference could have various causes. 

The first one, of observational nature, concerns systematic uncertainties on the 
$Li$ abundances. 
As said previously, the derivation of the lithium abundance in halo stars with
the high precision needed requires a fine knowledge of the physics of stellar 
atmosphere (effective temperature scale, population of different ionization states, 
non LTE effects at 1D and further on at 3D, Asplund et al. 2003).
However, the 3D, NLTE abundances are very similar to the 1D, LTE results,
 but, nevertheless, 3D models are now compulsory to extract lithium abundance from
 poor metal halo stars (see also Barklem et al. 2003).

Secondly, modification of the surface abundance of $Li$ by nuclear burning
all along the stellar evolution is discussed for a long time in the
literature. 
There is no lack of phenomena to disturb the $Li$ abundance
(rotational induced mixing, mass loss, see Theado and Vauclair 2001, and
Pinsonneault et al. 2002). 
However, the flatness of the plateau over three decades in metallicity and the 
relatively small dispersion of data represents a real challenge to stellar modeling. 
New data on $^{6}Li$ in halo stars are eagerly awaited since they will constrain more 
severely the potential destruction of \sep\ (see Vangioni-Flam et al. 1999). 
Finally, even taking into account the $Li$ upper limit 
of the Bonifacio et al. evaluation, the inconsistency persists.

The origin of the discrepancy between the WMAP+SBBN $Li/H$ calculated value
and that deduced from halo stars observations remains a challenging issue.
Large systematic errors on the 12 main nuclear cross sections are 
excluded (DAA) so that new physics has to be invoked if large observational bias can be themselves excluded 
.    
Both SBBN and CMB models use the minimal number of potential parameters 
(even a single one for SBBN) so that their extensions can be considered.
For instance, recent theories that could affect BBN include the time variation of 
 coupling constants (Ichikawa \& Kawasaki, 2002), 
the modification of 
the expansion rate during BBN induced by quintessence (Salati 2002), modified gravity
(Serna et al. 2002)  or neutrino degeneracy (Orito et al. 2002).  
These are fundamental issues on which BBN and CMB analyses could shed light.

However, first of all, the influence of {\it all} nuclear reactions needs to be 
evaluated before any conclusion.

\section{Nuclear uncertainties}
\label{s:reac}

The Monte--Carlo calculations using the DAA rate uncertainties introduced 
above provide the global uncertainties on yields. 
Here we present the effect of individual rate uncertainties
for the main reactions (DAA) but also for other reactions that have been, 
up to now, neglected.
It is well known that the valley shaped curve representing $Li/H$ as a 
function of $\eta$ is due to two modes of \sep\ production.
One, at low $\eta$ produces \sep\ directly via $^3$H($\alpha,\gamma)^7$Li
while \sep\ destruction comes from $^7$Li(p,$\alpha)^4$He.
The other one, at high  $\eta$, leads to the formation of $^7$Be 
 through $^3$He($\alpha,\gamma)^7$Be while
$^7$Be destruction by $^7$Be(n,p)$^7$Li is inefficient because of the 
lower neutron abundance at high density; ($^7Be$ later decays to \sep).
Since the WMAP results point toward the high $\eta$ region, we will pay 
a peculiar attention to $^7$Be synthesis.

\begin{figure}[htb]
\includegraphics[width=8cm]{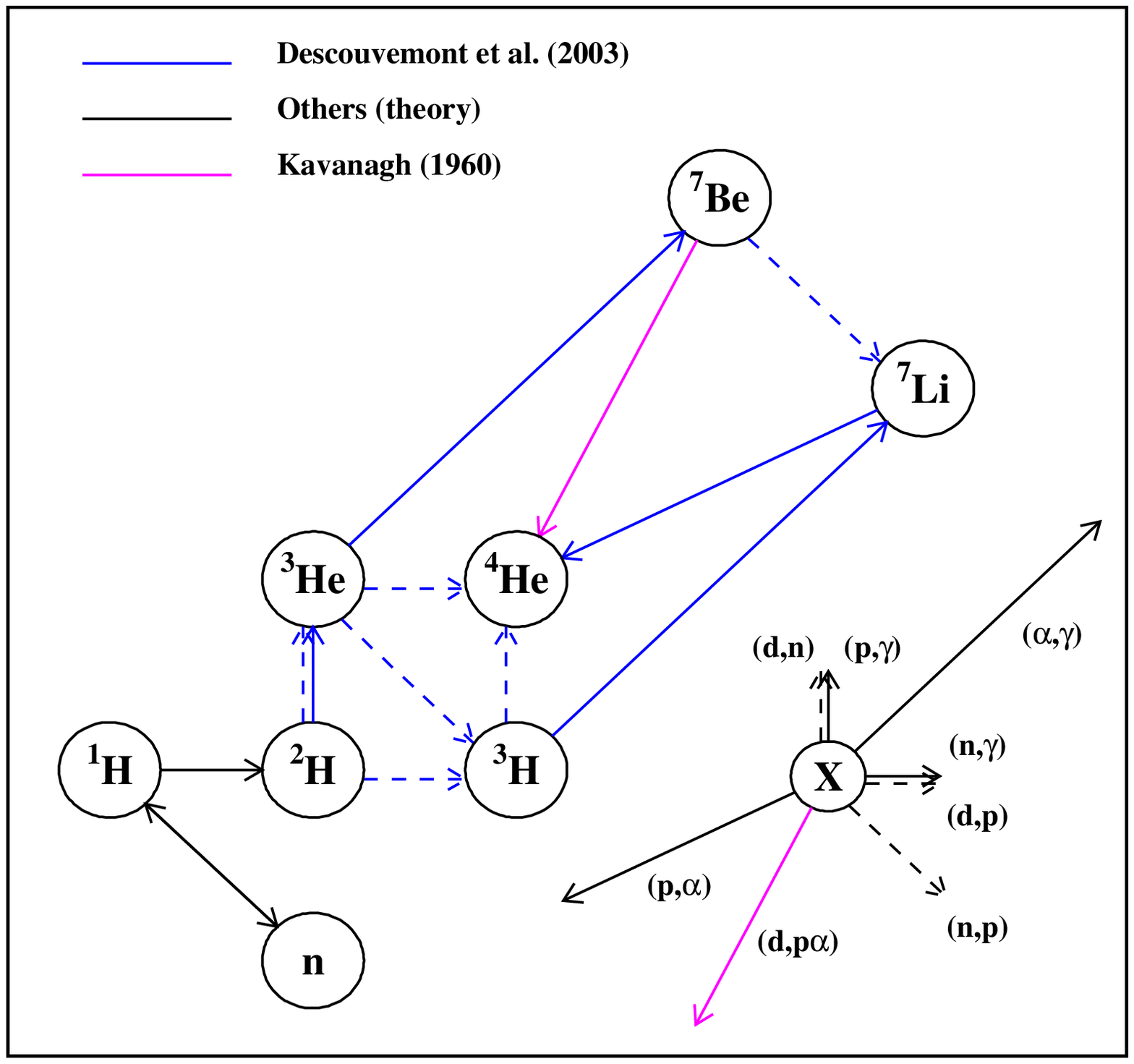}
\caption{The 12 main SBBN reactions plus $^7$Be(d,p)2$^4$He.}
\label{f:bbn}
\end{figure}

In Table~\ref{t:brux} are represented the maximum uncertainties on \hli\
isotopes arising from the rates of the 10 main nuclear reaction involved 
in SBBN using the results of DAA.
More precisely, $X_H$ (respectively $X_L$) represents the mass fraction
of a given isotope when one of the reaction rate is set to its $+1\sigma$ limit
(respectively $-1\sigma$ limit) and the {\em maxima} of the quantities $X_H-X_L$ 
for \qua\ and $\log\left(X_H/X_L\right)$ [i.e. dex]
for the other isotopes.
By {\em maximum}, we mean the value
having the maximum absolute value when $\eta$ spans the range between 
10$^{-10}$ and 10$^{-9}$.
Variations lower than 0.01 dex (10$^{-3}$ for $Y_P$) are not shown.  
From this table, we see that the reactions whose uncertainties affect most
\sep\ are $^2$H(p,$\gamma)^3$He, $^3$H($\alpha,\gamma)^7$Li,  
$^7$Li(p,$\alpha)^4$He for the low $\eta$ region and 
$^3$He($\alpha,\gamma)^7$Be for the (high $\eta$) region of interest.

\begin{table}
\caption{Influential reactions and their sensitivity to nuclear uncertainties
for the production of \qua, \deu, \tro\ and \sep\ in SBBN.}
\begin{flushleft}
\begin{tabular}{l|c|c|c|c}
\hline
Reactions& $^4$He&D&$^3$He&$^7$Li\\
\hline
& $(X_H-X_L)_{\mathrm max}$ & 
\multicolumn{3}{c}{$(\log\left(X_H/X_L\right))_{\mathrm max}$}\\
\hline
$^2$H(p,$\gamma)^3$He&-.-&-0.030&0.022&0.034\\
\hline
$^2$H(d,n)$^3$He&-.-&-0.009&0.007&0.011\\
\hline
$^2$H(d,p)$^3$H&-.-&-0.008&-0.008&0.003\\
\hline
$^3$H(d,n)$^4$He&-.-&-.-&-0.003&-0.004\\
\hline
$^3$H($\alpha,\gamma)^7$Li&-.-&-.-&-.-&0.038\\
\hline
$^3$He(d,p)$^4$H&0.0022&-.-&-0.018&-0.017\\
\hline
$^3$He(n,p)$^3$He&-.-&-.-&-0.006&-0.004\\
\hline
$^3$He($\alpha,\gamma)^7$Be&-.-&-.-&-.-&0.049\\
\hline
$^7$Li(p,$\alpha)^4$He&-.-&-.-&-.-&-0.039\\
\hline
$^7$Be(n,p)$^7$Li&-.-&-.-&-.-&-0.003\\
\hline
\end{tabular}
\end{flushleft}
\label{t:brux}
\end{table}

Since we are now interested in the precise determination of the isotopic
yields, it is important to check that besides the 12 main reactions of SBBN
the remaining ones are sufficiently known and do not induce any further
uncertainties. 
 
Rather than estimating the uncertainties on tens of reaction rates and
calculating the corresponding uncertainties on yields, we calculated the
yield variations when the rates are scaled by arbitrary factors.
If a variation of a reaction rate induces a significant change in the yield,
it will be the signal that this reaction should be studied in closer detail
and that the rate uncertainty should be calculated.
This is based on the prejudice that most of the reactions between A=1 and
A=12 have a negligible influence on isotope yields and hence that they need
not be known precisely.
To do so, we allowed the rates of the 43 reactions between 
$^2$H(n,$\gamma)^3$H 
and $^{11}$C(p,$\gamma)^{12}$N, whose rate uncertainties are not documented 
to vary by factors of 10, 100 and 1000 above their nominal 
rate and calculated the corresponding variation on the $^4$He, D, 
$^3$He, $^7$Li yields.
(Since the contribution of these reactions to these four isotopes is already 
considered negligible, it is irrelevant to consider lower rates.)
In many cases these factors may be excessive because the
rates are based on analysis of existing experimental data or on theory.
However, one should note for instance that in the new NACRE
compilation (Angulo et al. 1999) several rates differ from the previous 
ones (CF88) by several orders of magnitude. 
This is the case, in particular, of the $^{10}$B(p,$\alpha)^{7}$Be reaction 
whose rate has drastically changed between CF88 and NACRE because of new 
experimental data (Angulo et al. 1993). 
This has lead to a change of a factor of $\approx$10 in the SBBN
\dix\ yield (Vangioni--Flam et al., 2000).
In addition, several rates come from
estimates that have not been revisited for more than 30 years and could be
wrong or obsolete by unpredictable factors. 
This is might happen, in particular, for reactions involving unstable nuclei.
For instance, in another context, the $^{18}$F(p,$\alpha)$ reaction rate 
remains uncertain by several orders of magnitude, even at a few 10$^8$~K
(Coc et al., 2000).
So in a first step, we use these arbitrary variations, in many cases excessive, 
to select the most influential reaction rates. 
In that way we can eliminate from a more detailed study the many reactions 
whose influence remain negligible even if their rate is increased by a 
factor as large as 1000.
Then, in a second step, having drastically reduced the number of reactions, 
we discuss their actual nuclear uncertainties.

Table~\ref{t:vari} lists the few reactions, for which a 
variation of their rates by up to an arbitrary factor of 1000 induces a 
variation of the yields by more than 0.01~dex for \hli.
It shows that there are only four reactions that can lead to a factor of at 
least 3 (0.5 dex) on \sep\ yield when their rates are artificially 
increased by up to a factor of 1000 : $^3$H(p,$\gamma)^4$He, 
$^4$He($\alpha$,n)$^7$Be, $^7$Li(d,n)2$^4$He and $^7$Be(d,p)2$^4$He. 
It remains to check if such a huge increase in these reaction rates is
possible. As we will see, this is generally ruled out by existing data.

A factor of $\approx$1000 increases of the $^3$H(p,$\gamma)^4$He rate would be 
needed 
to reduce the \sep\ yield by a factor of 3.
This is excluded because, since CF88, this reaction cross section has been 
measured precisely by Hahn et al. (1995) and Canon et al. (2002) over the 
BBN energy range. 
The small changes in $S$-factor brought by these experiments (e.g. a 
$\approx$40\%  {\it reduction} relative to CF88 at a Gamow peak energy
corresponding to $T_9$ = 1) rule out any possible influence in BBN.
In any case, as seen in Fig.~\ref{f:bbn}, this reaction could only affect
the low baryonic density branch, $^3$H($\alpha,\gamma)^7$Li, and not
the WMAP density region.  

The reaction rate for $^7$Li(d,n)2$^4$He comes from an analysis by Boyd 
et al. (1993) of \sep\ destruction in BBN. 
A factor of 100 increase could reduce the \sep\ production by a factor of 
$\approx$3.
Even though, no rate uncertainties are provided by Boyd et al., this seems
quite unlikely as their analysis is based on experimental data available
in the BBN energy range. 
Nevertheless, as for the previous reaction this could only influence the direct
\sep\ formation i.e. the low baryonic density region.

On the contrary, the $^4$He($\alpha$,n)$^7$Be reaction (Q=-18.99~MeV) could 
affect \sep\ production at high $\eta$, where it is formed as $^7$Be
(Fig.~\ref{f:bbn}), and through $^7$Be destruction by the reverse reaction 
$^7$Be(n,$\alpha\gamma$)$^4$He.
However, the rate of this latter is negligible compared to the main destruction mechanism :
$^7$Be(n,p)$^7$Li (Fig.~\ref{f:bbn}) where a $\ell$=0 resonance dominates  
while $\ell$=0 is forbidden in $^7$Be(n,$\alpha\gamma$)$^4$He due to the 
symmetry of the outgoing channel.

\begin{figure}[htb]
\includegraphics[width=8cm]{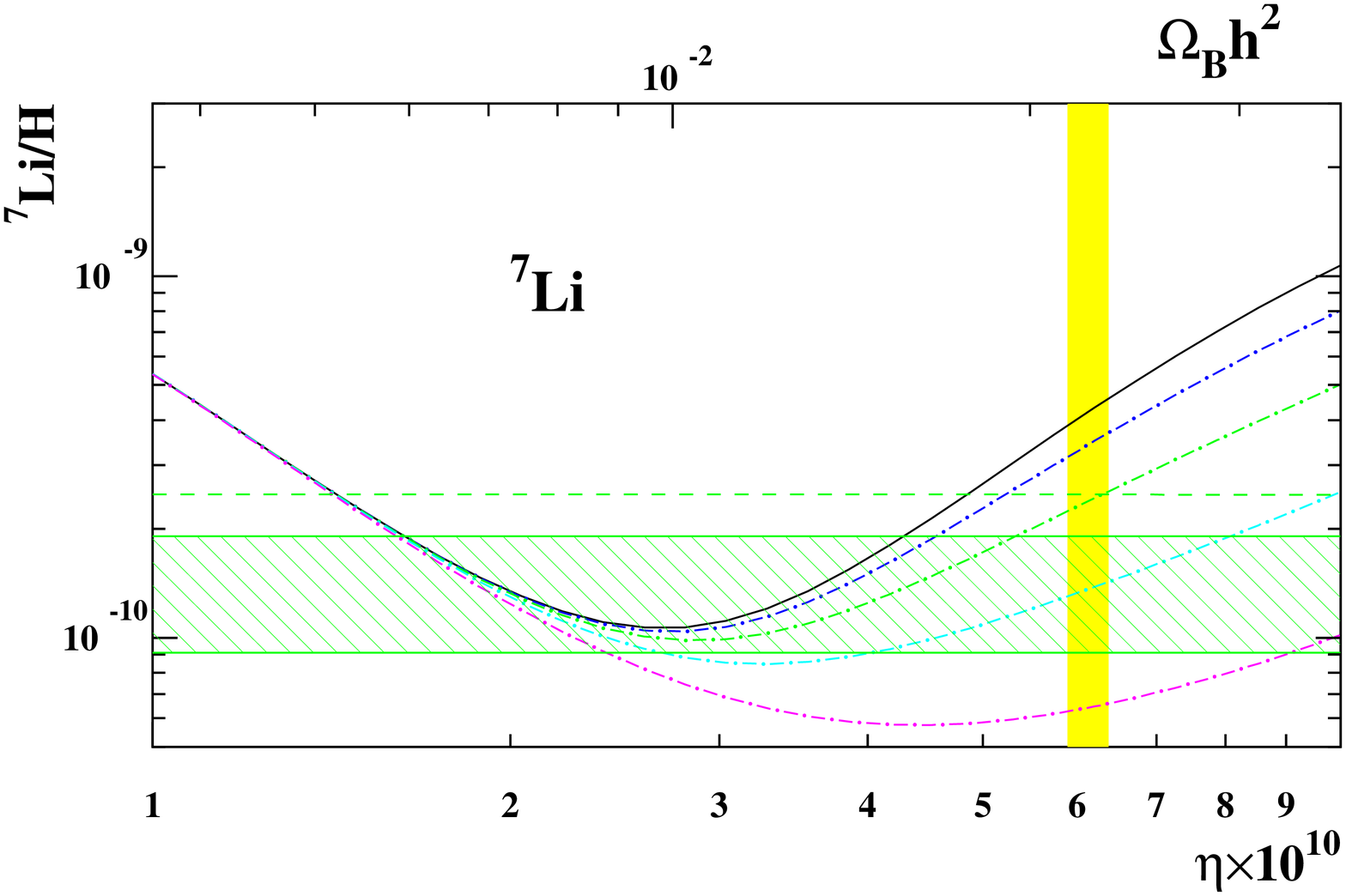}
\caption{Same as Figure~\ref{f:heli}, lower panel, but including the
effect of $^7$Be(d,p)2$^4$He rate variations while other reaction rates are
set to their nominal values. 
The solid curve is the reference where the $^7$Be(d,p)2$^4$He rate from CF88 is 
used, while the dash--dotted curves correspond to an increase of the rate by factors 
of 30, 100, 300 and 1000.}
\label{f:be7dp7}
\end{figure}

\begin{figure}[htb]
\includegraphics[width=8cm]{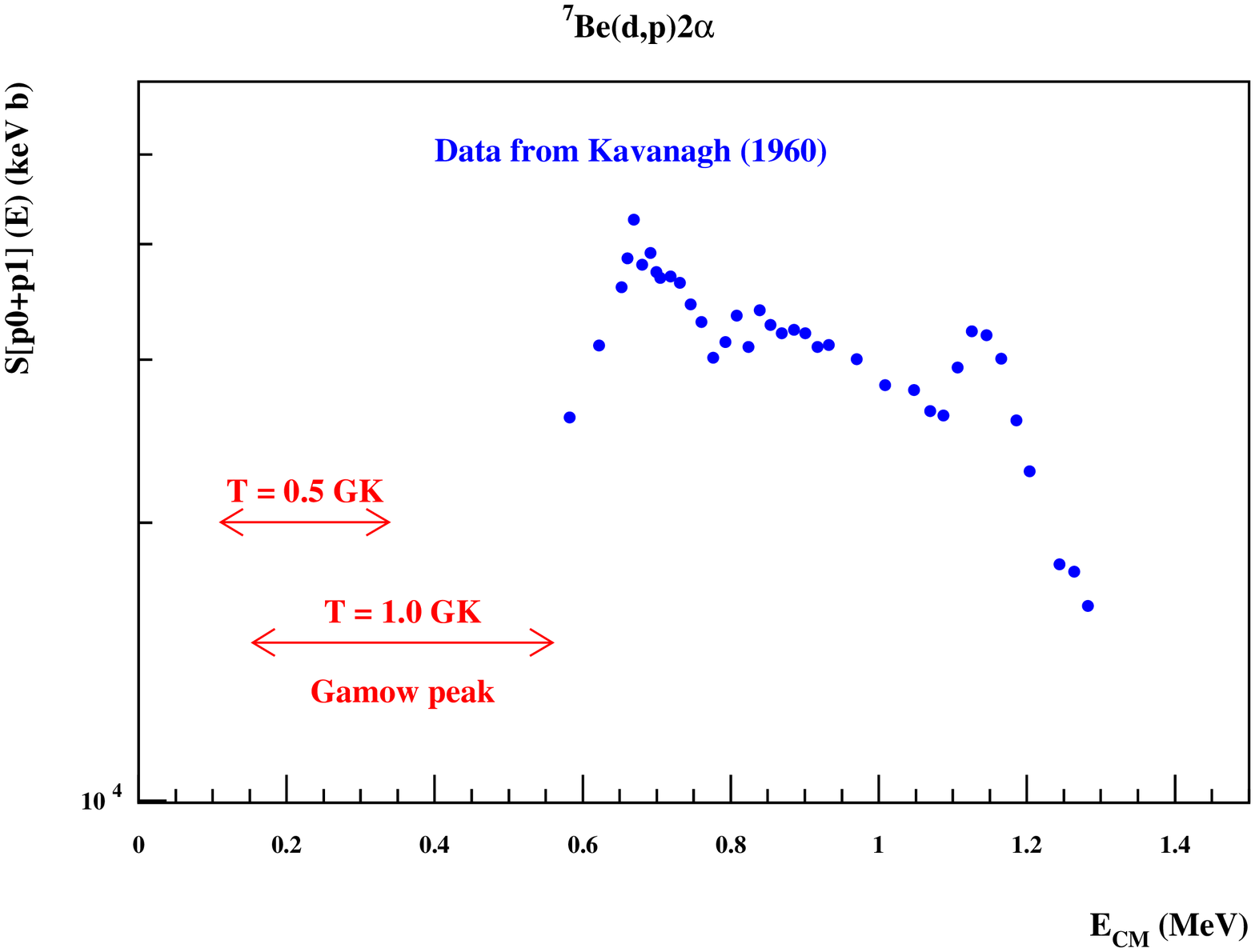}
\caption{The only experimental data available for the $^7$Be(d,p)2$^4$H
reaction from Kavanagh (1960). The displayed $S$--factor is calculated 
as in Parker (1972) from the differential cross section at 90$^\circ$  
($\times4\pi$) leading to the ground and first $^8$Be excited states.  
Note that no data is available at SBBN 
energies as shown by the Gamow peaks for T$_9$ = 1 and 0.5.}
\label{f:be7dpk}
\end{figure}

The last reaction in Table~\ref{t:vari}, $^7$Be(d,p)$^8$Be($\alpha)^4$He
is then the most promising in view of reducing the discrepancy between 
SBBN, \sep\ and CMB observations.
$^7$Be+d could be an alternative to $^7$Be(n,p)$^7$Li for the 
destruction of $^7Be$ (see Fig.~\ref{f:bbn}), 
by compensating the scarcity of neutrons at high $\eta$.
Figure~\ref{f:be7dp7} shows the effect of an increase of the 
$^7$Be(d,p)2$^4$He reaction rate: a factor of $\gap$ 100 
could alleviate the discrepancy.
The rate for this reaction (CF88) can be traced to an estimate by Parker (1972) 
who assumed for the astrophysical $S$--factor a constant value of 10$^5$~kev.barn. 
This is based on the single experimental data available (Kavanagh, 1960).
To derive this $S$--factor, Parker used the measured differential cross 
section at 90$^\circ$ and assumed isotropy of the cross section. 
Since Kavanagh measured only the p$_0$ and p$_1$ protons 
(i.e. feeding the $^8$Be ground and first excited levels), Parker 
introduced an additional but arbitrary factor of 3 to take into account
the possible population of higher lying levels. Indeed, a level at 11.35~MeV
is also reported (Ajzenberg-Selove 1988). This factor should also include the 
contribution of another open channel in $^7$Be+d: $^7$Be(d,$\alpha)^5$Li
for which no data exist. 
The experimental data (Kavanagh, 1960) is displayed in 
Fig.~\ref{f:be7dpk} showing the two expected resonances at 0.7 and 1.2~MeV
(Ajzenberg-Selove, 1988). A third one at 0.6~MeV is excluded because of isospin 
selection rules.
\sep\ and $^7$Be Big Bang nucleosynthesis take place when the 
temperature has decreased below T$_9$=1. The Gamow peaks for
T$_9$=1 and 0.5 displayed in Fig.~\ref{f:be7dpk} show that
there are no experimental data at SBBN energies.
A seducing possibility to reconciliate, SBBN, \sep\ and CMB observations 
would then be that new experimental data below $E_d$ = 700~keV 
($E_{cm}\approx$0.5~MeV) 
for $^7$Be(d,p)2$^4$He [and $^7$Be(d,$\alpha)^5$Li]
would lead to a sudden increase in the $S$--factor 
as in $^{10}$B(p,$\alpha)^{7}$Be (NACRE).
This is not supported by known data, but considering 
the cosmological or astrophysical consequences, this is definitely an
issue to be investigated.  
Accordingly, an experimental study of this reaction
 will be performed soon at Louvain la Neuve.

\begin{table}
\caption{\label{t:fx}Test of yield sensitivity to reactions rate variations:
factor of 10,100,1000 (see text).}
\begin{tabular}{l|l|c|c|c|c}
\hline
Reaction
& Ref. &$^4$He&D&$^3$He&$^7$Li\\
\hline
&& $(X_H-X_L)_{\mathrm max}$ & 
\multicolumn{3}{c}{$(\log\left(X_H/X_L\right))_{\mathrm max}$}\\
\hline
$^2$H(n,$\gamma)^3$H& Wag69 &0.003&~~-.-~~&-.-&-.-\\ 
                          & & 0.025&-0.010&-.-&-0.011\\ 
&                           & 0.110&-0.073&-0.048&-0.078\\ 
\hline
$^3$H(p,$\gamma)^4$He& CF88 & -.-&-.-&0.012&0.074\\ 
&                           & 0.003&-0.017&0.055&0.26\\ 
&                           & 0.018&-0.058&0.14&-0.56\\ 
\hline
$^3$He(t,np)$^4$He& CF88 & -.-&-.-&-.-&-.-\\ 
&                        & -.-&-.-&-.-&-0.012\\ 
&                        & -.-&0.053&-0.026&-0.092\\ 
\hline
$^4$He($\alpha$,n)$^7$Be& Wag69 & -.-&-.-&-.-&-0.056\\ 
&                               &-.-&-.-&-.-&-0.36\\ 
&                               & -.-&-.-&-.-&-1.1\\ 
\hline
$^7$Li(d,n)2$^4$He& Boy93 &-.-&-.-&-.-&-0.10\\ 
&                         &-.-&-.-&-.-&-0.44\\ 
&                         &-.-&-.-&-.-&-1.1\\ 
\hline
$^7$Li(t,2n)2$^4$He& MF89 &-.-&-.-&-.-&-.-\\ 
&                         &-.-&-.-&-.-&-.-\\ 
&                         &-.-&-.-&-.-&-0.055\\ 
\hline
$^7$Be(d,p)2$^4$He& CF88 &-.-&-.-&-.-&-0.047\\ 
&                        &-.-&-.-&-.-&-0.34\\ 
&                        &-.-&-.-&-.-&-1.0\\ 
\hline
\end{tabular}
\label{t:vari}
Wag69: Wagoner 1969; CF88: Caughlan \& Fowler 1988;\\ 
Boy93: Boyd et al. 1993; MF89: Malaney \& Fowler 1989. 
\end{table}

\section{Conclusions}

In conclusion, the recent WMAP experiment has to be acknowledged as a great progress, specifically
concerning the evaluation of the baryon content of the Universe. This leads the nuclear 
astrophysicists to refine their calculations.  
We have improved SBBN calculations taking into account a new nuclear physics analysis (DAA) of 
SBBN reaction rates. 
The consistency between WMAP results and D/H data from the remote cosmological clouds on the 
line of sight of high redshift quasars strengthens the deduced baryonic density.
However, a significant discrepancy is observed for lithium. Nuclear effects, as 
in particular higher $^7$Be+d reaction rates (see above), could 
reconciliate 
calculations and  observations. If not, new and exciting  astrophysical or physical 
effects will have to be considered.

\acknowledgments

We warmly thank Roger Cayrel, Guillaume H\'ebrard, 
 and Fr\'ed\'eric Th\'evenin for fruitful discussions.
 We thank also Keith Olive for his permanent usefull collaboration. Finally thanks 
 very much to Martin Lemoine for reading the manuscript.
This work has been supported by the PICS number 1076 of INSU/CNRS.

\end{document}